# DISCOVERY OF GAMMA-RAY EMISSION FROM M31 VIA *FERMI* LAT

Ögelman, H.,[1,2,3] Aksaker[1], N., Anılan, S.[1], Dereli, H.[1] Emrahoğlu, N[1]., Yegingil, I.[1]

[1] Physics Department, Çukurova University, 01330 Adana, Turkey
[2] Sabancı University, Orhanlı-Tuzla, 34956 Istanbul, Turkey
[3] Physics Department, University of Wisconsin, Madison, WI 53706, USA

Abstract

2 years worth of archival FERMI-LAT data was used to search for the gamma-ray emission from the Andromeda galaxy. The data show no noticeable elliptical image. Subsequent on-off source aperture photometry analysis using a CO image template show a $7\sigma$ excess in the number of on-source apertures in comparison to the off-source apertures, yielding a flux of $(4.95\pm0.71)\times10^{-8}$ photons cm$^{-2}$ s$^{-1}$ for E>100 MeV.

**1 Introduction** Our local group of galaxies offer the only chance to verify what we have learned about the gamma-ray emission of our own galaxy the Milky Way, owing to the proximity of these objects (Vasiliki et.al 2001). So far this has only been possible for LMC due to its very close position (55kpc) (Abdo *et al*. 2010). In contrast another major member of our local group M31 Andromeda Galaxy has been elusive. *EGRET* has set an upper limit for it (Bloom *et al*.1999) :The $2\sigma$ upper limit for energies above 100 MeV was $1.6 \times 10^{-8}$ photons cm$^{-2}$ s$^{-1}$ The main reason why M31 has been so elusive is due to it's distance (0.812 Mpc) and the lack of a *FERMI* class observatory till now.

## 2 The data base used in this work

*FERMI*-LAT data was downloaded from the all-sky survey database as it was available near mid July, 2010. It covered the time span 08-2008 to 07-2010, about 2 years of observations. Photons with energy E > 100 MeV were extracted from within a radius of 10° centered on M31. After event class and zenith angle corrections we had 128,425 photons to work with. These events thus formed the basis of the following investigation. For the description of the LAT-detectors see Atwood *et al.* (2009). A celestial map constructed from these events is shown in Figure 1.

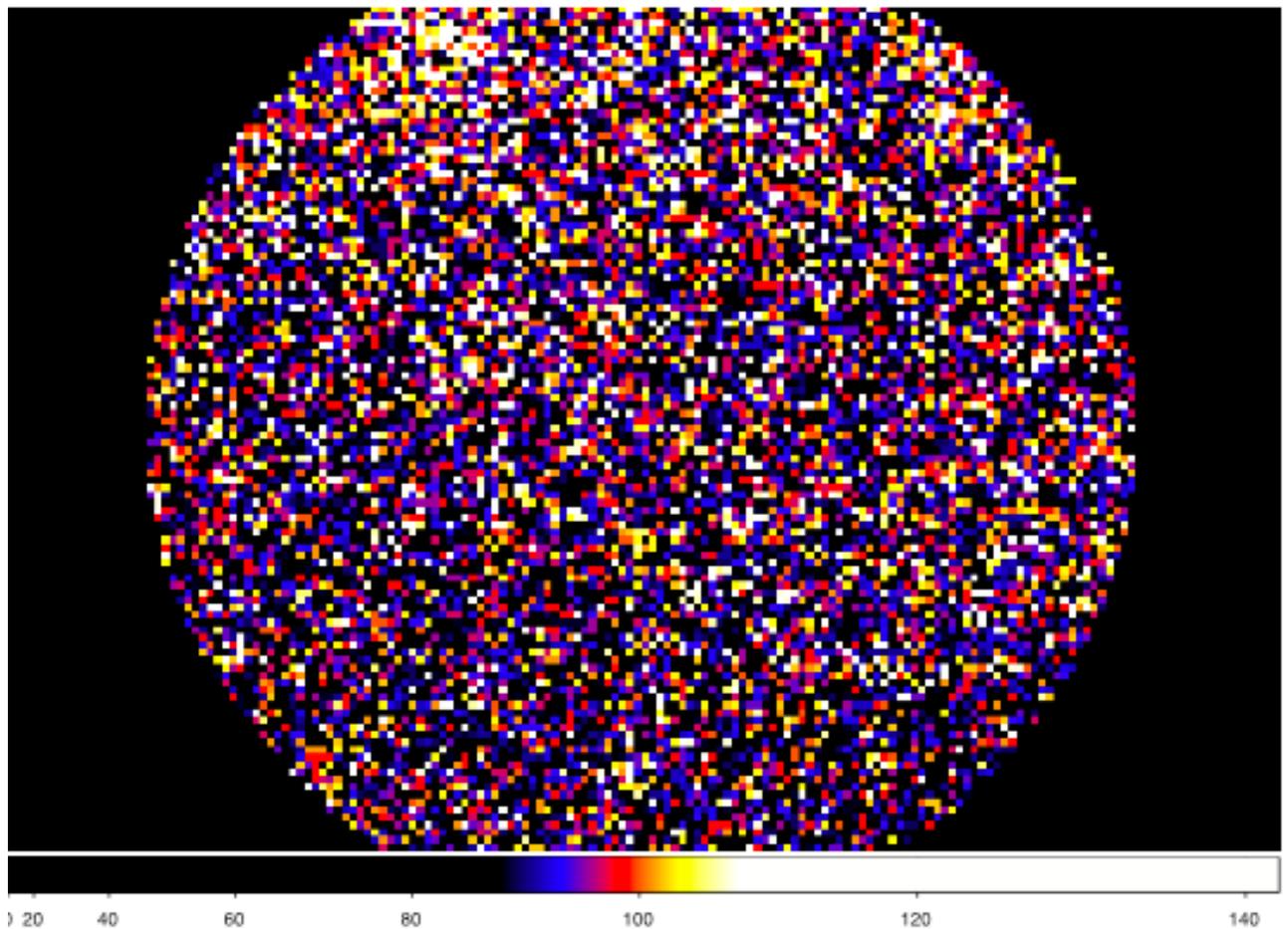

**Fig1**. Map of the M31 region for the downloaded Fermi-LAT data, centered on M31 extending to 10°.

The first impression from the figure is that there is no elliptical shaped image delineating the image of the M31 galaxy seen in other wavelengths, see for example Figure 2, a VLA image which has been downloaded from the NED data base.

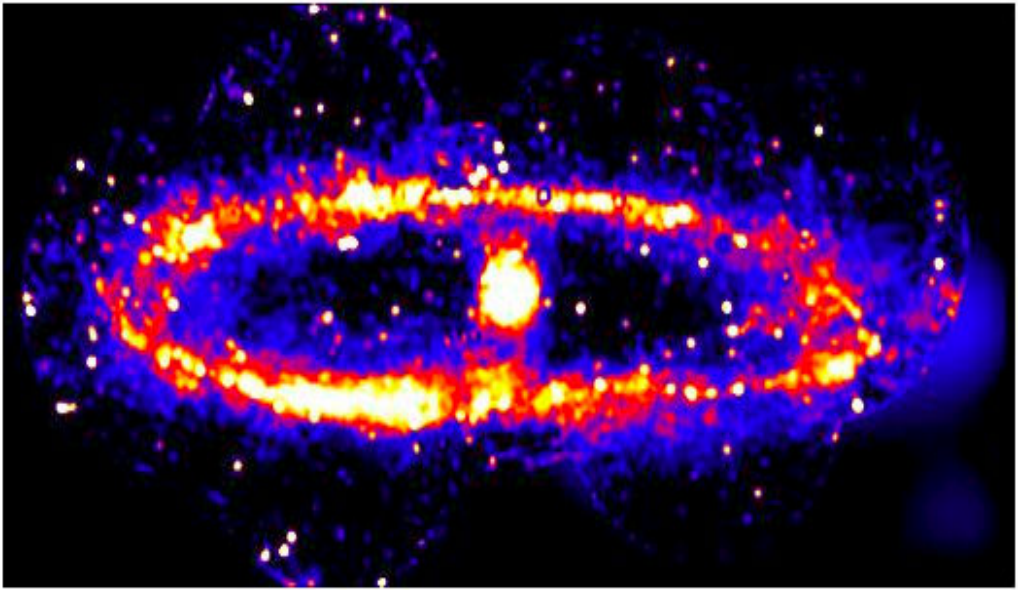

**Fig2** A VLA image of M31 showing the location of the energetic electrons and delineating the shape of the galaxy

Yet we expect that there should be a minimum level of γ-ray emission from M31 produced by the inelastic scattering of cosmic-rays (CR) with the ambient Interstellar Medium (ISM) through the reactions

$p_{CR} + p_{ISM} \rightarrow \pi^o + \pi^o \rightarrow \gamma + \gamma$ .

This process is the basis of the γ-ray emission of our Milky Way. Fig 3 is the *Fermi* image of the Milky Way showing this γ-ray emission.

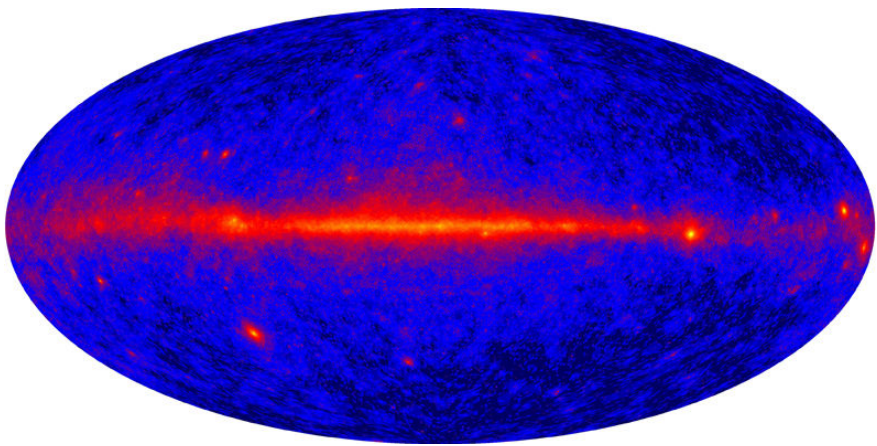

**Figure3** FERMI/LAT image of our Milk Way

## 3 Aperture Photometry

Since we know there is ISM gas in M31 and since we suspect strongly that there are cosmic-rays there, we develop a strategy to observe the $\pi^0$ decay gamma-rays. For the template of this emission we use the CO emission map downloaded from NED database. This map is shown in Figure 4

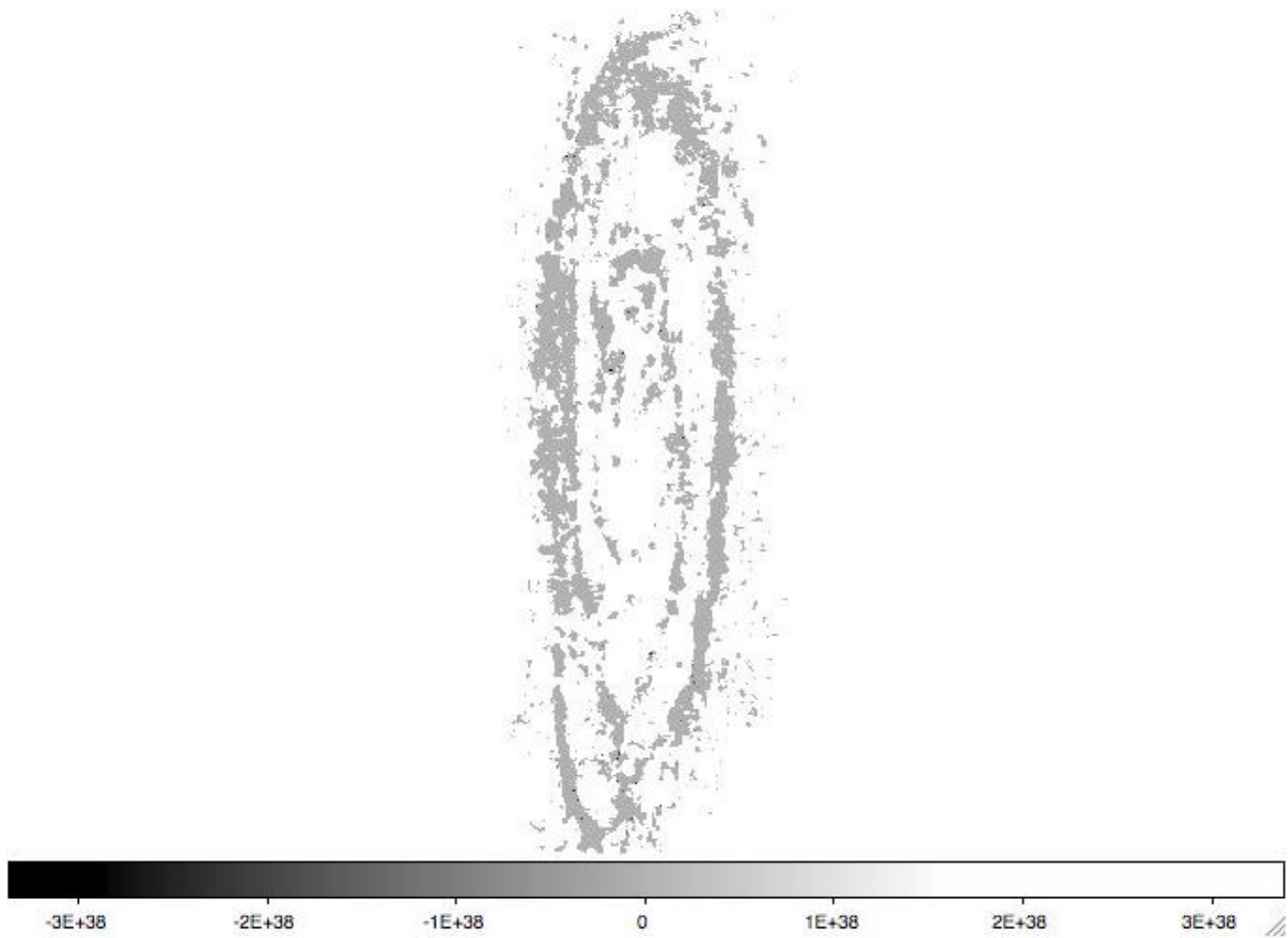

**Figure4** CO image of M31 downloaded from the NED database.

We selected 6 apertures whose central coordinates cover the long axis of the ellipse seen in the CO image. The radius of the apertures were chosen as $0.1°$, small enough to do the micro-aperture photometry and small enough to ensure statistical independence of the apertures; the psf of the counted photons would still be larger than these apertures. However, reducing this radius too much would end up getting no events in the aperture. Six apertures were chosen, the number limited by the completion of the ellipse coverage. The off-source background was chosen at the western-edge of the CO image at about the same declination as the on-source apertures. Their radius was also kept at $0.1°$ and their number was kept at 6 in order to preserve a good symmetry between source and background apertures.

## 4 Results of the aperture photometry

We first looked directly at the raw counts intersected by these apertures. The 6 on-source apertures yielded: 567+552+531+561+571+536 which added up to 3318 counts; the off-source results were: 551+513+213+489+511+497 adding up to 2774 counts; notice that none of the aperture counts were excessively large or small compared to the average, confirming that we had not selected any aperture on untypical regions. The net source -background counts are: 3318-2774=544 ±78.This corresponds to a detection significance of 7σ. Encouraged by this positive result, we tried to see something on the image by Gaussian smoothing of the image on Fig1. The result is shown in Figure 5.

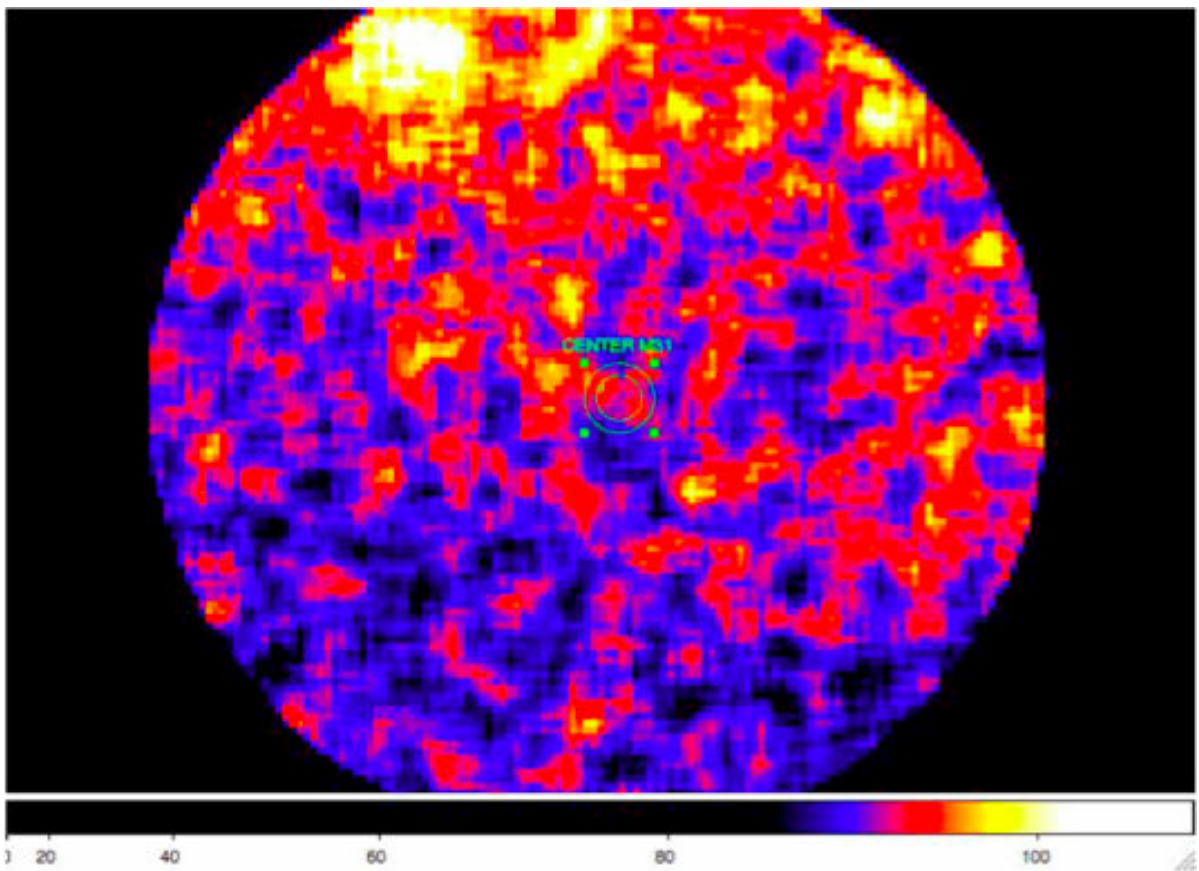

**Fig 5** Gaussian smoothed version of Figure 1

## 5 Discussion

**5.1 Lack of point-sources:** Is the lack of point-sources on the image surprising? The answer would be No if we take the brightest source in our galaxy, the Crab Nebula and put it at the distance of M31. FERMI count rate for E>100 MeV would be $10^{-21}$ cts cm$^{-2}$ s$^{-1}$, a very low flux even for FERMI.

**5.2 Consequences of the measured flux:**

The exposure calculated from the aperture photometry as $1.1 \times 10^{10}$ cm$^2$-s$^{-1}$, giving a measured flux of $(4.945 \pm .71) \times 10^{-8}$ photons cm$^{-2}$ s$^{-1}$. The 2σ upper-limit from EGRET is $1.6 \times 10^{-8}$ photons cm$^{-2}$ s$^{-1}$ for E>100 MeV (Bloom *et al.* 1999). Within 3σ limits the two results would be in agreement. Our measured flux implies a photon luminosity for M31 of $10^{42}$ photons s$^{-1}$. If we correct for the total aperture for M31 the luminosity will increase to $(6.0 \pm 0.9) \times 10^{42}$ photons s$^{-1}$ for E>100 MeV.

**5.3 Comparison with the Milky Way**

The luminosity of our Milky Way due to cosmic-ray ISM inelastic collisions has been estimated as $3 \times 10^{41}$ photons s$^{-1}$. Thus:
☐ M31 is 20 times more luminous than Milky Way,
☐ Since it is ≈10 times more massive it can readily have 10 x the ISM interstellar mass of the Milky Way,
☐ and, if the cosmic-ray flux at M31 is twice that at the Milky Way, we can easily account for the excess by a factor of 20.

A more reasonable scaling may go something like this:
i) if the cosmic-rays originate within a galaxy, the CR production rate ∝ supernova rate ∝ stellar mass

ii) number of target protons ∝ ISM mass ∝ stellar mass.

Hence the γ-ray luminosity ∝ (stellar mass)$^2$. In case cosmic-rays are extragalactic then we would expect the γ -ray luminosity ∝ (stellar mass)$^1$. More generally, if the cosmic rays have both intra- galactic supernova and extragalactic origins, the γ-ray luminosity may be described by some other power of the galaxy mass. For the factor of 20 larger γ -ray luminosity of M31, compared to that of the Milky Way while there is only a factor of 10 between the masses of the two galaxies, the power–law exponent needs to be 1.3, which appears to favor an extragalactic origin cosmic-rays. Further data should be processed before reaching conclusions about cosmic-ray models.

**Acknowledgement** H. Ö. thanks colleagues at Sabanci University & Cukurova University for the hospitality during his visit. The main ideas in this paper have been proposed as a FERMI proposal to NASA which was not accepted.